# A Dynamical Model Reveals Gene Co-Localizations in Nucleus

Jing Kang[1,2]◊, Bing Xu[3]◊, Ye Yao[3], Wei Lin[3], Conor Hennessy[1], Peter Fraser[1]*, Jianfeng Feng[3,2]*

1 Nuclear Dynamics Laboratory, The Babraham Institute, Cambridge, United Kingdom, 2 Centre for Scientific Computing, Warwick University, Coventry, United Kingdom, 3 Centre for Computational Systems Biology, Fudan University, Shanghai, People's Republic of China

**Abstract**

Co-localization of networks of genes in the nucleus is thought to play an important role in determining gene expression patterns. Based upon experimental data, we built a dynamical model to test whether pure diffusion could account for the observed co-localization of genes within a defined subnuclear region. A simple standard Brownian motion model in two and three dimensions shows that preferential co-localization is possible for co-regulated genes without any direct interaction, and suggests the occurrence may be due to a limitation in the number of available transcription factors. Experimental data of chromatin movements demonstrates that fractional rather than standard Brownian motion is more appropriate to model gene mobilizations, and we tested our dynamical model against recent static experimental data, using a sub-diffusion process by which the genes tend to colocalize more easily. Moreover, in order to compare our model with recently obtained experimental data, we studied the association level between genes and factors, and presented data supporting the validation of this dynamic model. As further applications of our model, we applied it to test against more biological observations. We found that increasing transcription factor number, rather than factory number and nucleus size, might be the reason for decreasing gene co-localization. In the scenario of frequency- or amplitude-modulation of transcription factors, our model predicted that frequency-modulation may increase the co-localization between its targeted genes.

Funding: This work is supported by BION (EU) and CARMEN (EPSRC, UK) grants to Jianfeng Feng, and the Biotechnology and Biological Science Research Council and the Medical Researh Council, UK, to Peter Fraser. The funders had no role in study design, data collection and analysis, decision to publish, or preparation of the manuscript.

Competing Interests: The authors have declared that no competing interests exist.

* E-mail: peter.fraser@bbsrc.ac.uk (PF); jianfeng.feng@warwick.ac.uk (JF)

◊ These authors contributed equally to this work.

## Introduction

A central theme in the regulation of transcription is the binding of transcription factor proteins to specific sites along the DNA. Though these sites can be several tens or hundreds of kilobases from a target gene promoter, regulation is achieved by the formation of chromatin loops that bring the sites together to form transcriptional hubs. It is thought that proximity between distal regulatory elements and their target genes increases the local concentration of specific regulatory factors to affect transcriptional control. Recent studies have also shown that active genes co-localize in the nuclear space at focal concentrations of the active form of RNA Polymerase II (RNAPII) called transcription factories [1,2,3,4,5,6]. A genome-wide enhanced 4C (e4C) screen demonstrated that specific combinations of genes from different chromosomes share factories with a high frequency, suggesting that active genes have preferred transcription partners. Co-localization of these spatial gene networks at transcription factories was found to be dependent on the transcription factor Klf1, which co-regulates many of the partners [7]. Just as distal regulatory elements are thought to affect gene regulation by spatial clustering, intra- and inter-chromosomal associations between co-regulated genes may affect expression by creating specialized microenvironments that are optimized for their transcription. Thus, the transcriptional program of a cell may be reflected by, or may even be dependent upon, the spatial organization of the genome. The appreciation that a very large proportion of the genome is transcribed with relatively few transcription sites suggests that the organization of the transcriptional machinery plays a major role in shaping the nuclear organization of the genome. The positioning of genes, regulatory sequences and transcription factors in relation to each other and to landmarks in the nucleus, such as nuclear bodies and lamina, are important determinants in gene expression [8].

How specific subgroups of active genes and transcription factors come to be positioned at factories is still unknown. Gaining an understanding of the emergence of complex spatiotemporal patterns of behavior from the interactions between genes in a regulatory network poses a huge scientific challenge with potentially high industrial pay-offs [4,9,10,11,12]. Experimental techniques to dissect regulatory interactions on the molecular level are critical to this end. In addition to experimental tools, mathematical modeling and computer tools will be indispensable. As most genetic regulatory systems of interest involve many genes connected through interlocking feedback loops, an intuitive understanding of their behavior is hard to obtain. By explicating hypotheses on the topology of a regulatory network in the form of a computer model, the behavior of possibly large and complex

## Author Summary

Transcription is a fundamental step in gene expression, yet it remains poorly understood at cellular level. Textbooks are full of descriptions of promoter-bound transcription factors recruiting RNA polymerase, which initiates transcription before sliding along the transcription unit. However, increasing evidence supports the view that the DNA template bound with transcription factors slides through a relatively immobile RNA polymerase at discrete nuclear sites (known as transcription factories), rather than RNA polymerase sliding along DNA template. Based on this transcription factory model, we build a virtual space in which genes and transcription factors move randomly while transcription factories are immobile. We find that under a large number of parameter ranges, this simple dynamical model is valid for a number of experimental observations. Moreover, we suggest the occurrence of gene co-localization might be mainly due to limited numbers of transcription factors, rather than other factors such as nucleus size or transcription factory number. This work offers insight into the general principles of regulation of transcription and gene expression by simulating the translocation of transcriptional units (genes and transcription factors) using purely random diffusion processes that result in non-random organization of co-regulated genes.

regulatory systems can be predicted and explained in a systematic way. One of such recent examples is described in Misteli [13] and Rajapakse et al [14], where the authors developed a model based upon self-organization. It is probably the most successful model in the area, as confirmed in Misteli [13]. However, the model is phenomenological with an oversimplified system of Kuramoto oscillators and the random effect is largely ignored.

Here we have developed a model based upon known experimental data, aiming to account for experimental results and for predicting and guiding further experiments. Co-localization ratio is introduced to characterize the gene co-localizations in transcription factories. Within a wide parameter region, we demonstrate that gene co-localization is plausible for both two and three dimensional cases. Experimental data tells us that sub-diffusion is observed in various cell-cycle phases (S and G phase) in yeast [15], which implies that fractional Brownian might be required to model gene movement, at least locally in time and space. Using fractional Brownian motion, gene and gene pairs co-localized with transcription factories are estimated, and tested against experimental data obtained from RNA-immuno-FISH experiments. We find that the model, albeit simple, can account for observed experimental data.

Previous research [1] showed that mouse embryonic fibroblasts (MEFs) which have flattened nuclei have relatively high numbers of transcription factories (~2,000) while embryonic, fetal and adult erythroid cells and normal adult spleen, adult thymus, and fetal brain cells with spherical nuclei, have fewer transcription factories (100–300). Therefore, one may wonder what effects varying numbers of transcription factories and nuclear shape have on gene co-localization. Our simulation using both flattened and spherical nucleus, showed that co-localization is not very sensitive to the number of transcription factories, but is sensitive to the number of transcription factors.

Transcription factor entry to the nucleus may occur in two ways: either in a frequency-modulation mode (NF-κB for example [16]) or via amplitude modulation (Klf1 might be an example) fashion. Recently Cai et al. observed that Crz1, a stress-response transcription factor, translocates to nucleus in response to extracellular calcium signal, showing short bursts of nuclear localization [17] (frequency modulation). They proposed that frequency-modulation, rather than amplitude-modulation, of localization bursts of transcription factors may be a control strategy to coordinate gene responses to external signals. Interestingly, we found that frequency-modulation, in comparison with amplitude-modulation, facilitates gene co-localization. This might reveal a key advantage of frequency modulation over amplitude-modulation in coordinating gene expression in cell nuclei.

## Results

### Co-localization between genes

Recent studies show that active genes dynamically co-localize to shared transcription sites and that specific networks of genes share factories at high frequencies [1,7,18,19]. We built a model by randomizing the movement of genes and transcription factors, with a defined number of immobile transcription factories [6,18]. Live cell studies have shown that chromatin is highly mobile but regionally constrained within eukaryotic nuclei. We therefore created a defined space for random diffusion of genes (genes restricted to a square in the 2D model or a cubic in the 3D model) based on the observed mobility of chromatin *in vivo*. Each gene regulatory element is regarded as a point for simplicity, rather than as a polymer. The transcription process is simulated as follows: when a gene and its transcription factor come within a defined proximity, they bind and diffuse together for a limited time (which is called binding time). If the bound complex encounters a transcription factory before their binding time elapses, the gene engages with the polyerase and becomes active, remaining associated with the transcription factory until termination (the transcription time); if the gene-factor complex does not encounter a transcription factory during the binding time, the gene and factor dissociate and continue their brownian motion separately. During a productive transcriptional event, the transcription factor may be released from the gene before it finishes transcription and is available to randomly interact and bind another gene of the same family.

We examined the behavior between genes of the same family (X genes) and two different families of genes (X genes and Y genes) (see Fig. 1A). Genes and their corresponding factors (X factors and Y factors) are allowed to move randomly within the restricted region. However, to simulate the constrained diffusion of chromatin, the genes cannot exit the defined space. Transcription factors are allowed to exit the space, but exit of one factor is followed by entry of an identical factor on the opposite/same side of the space (We have simulated both cases and found no significant difference in the simulation). This maintains the concentration of factors within the space and simulates the observed behavior of factors to explore the entire nucleus. The number of genes and factors for each family are equivalent. X factors can only bind to X genes and Y factors to Y genes, and the two families of genes and factors are independent of each other. We also assume that each gene has only two states (Fig. 1B and 1C) — either it is being transcribed (X gene: $u(t)=1$; Y gene: $v(t)=1$) or not being transcribed (X gene: $u(t)=0$; Y gene: $v(t)=0$). Fig. 1B and 1C illustrate X-X gene co-localization and X-Y gene co-localization, respectively, where $u(t)$ is the state function of X gene and $v(t)$ the state function of Y gene.

The choices of parameter values are based on the literatures and previous experimental observations, and are presented in Table 1. For example in our 3D model, the number of genes, factors and factories in our restricted region (about 1/50 in terms of volume) is proportional to the total number of active genes in nucleus (10,000–20,000 alleles, around 2000 transcription factors, and 100–300 factories observed per nucleus, reviewed by [20]). The size of genes, transcription factors and factories were also chosen in consistency with biological data. Since transcription factor binding sites are often clustered in regulatory elements in chromatin, we have given genes a binding radius of 25 nm. The diffusion rate of a gene (0.001 $\mu m^2$/sec) and transcription factor (0.01 $\mu m^2$/sec) is chosen in consistency with previous work [20,21]. The volume of the restricted region is based on Chubb et al's results (displacements of genes were in the range between 0.5 $\mu m$ to 2 $\mu m$ [22]). The gene-factor binding time is on the timescale of seconds [23], and transcription time is consistent with the speed of the polymerase across a relatively small gene [24].

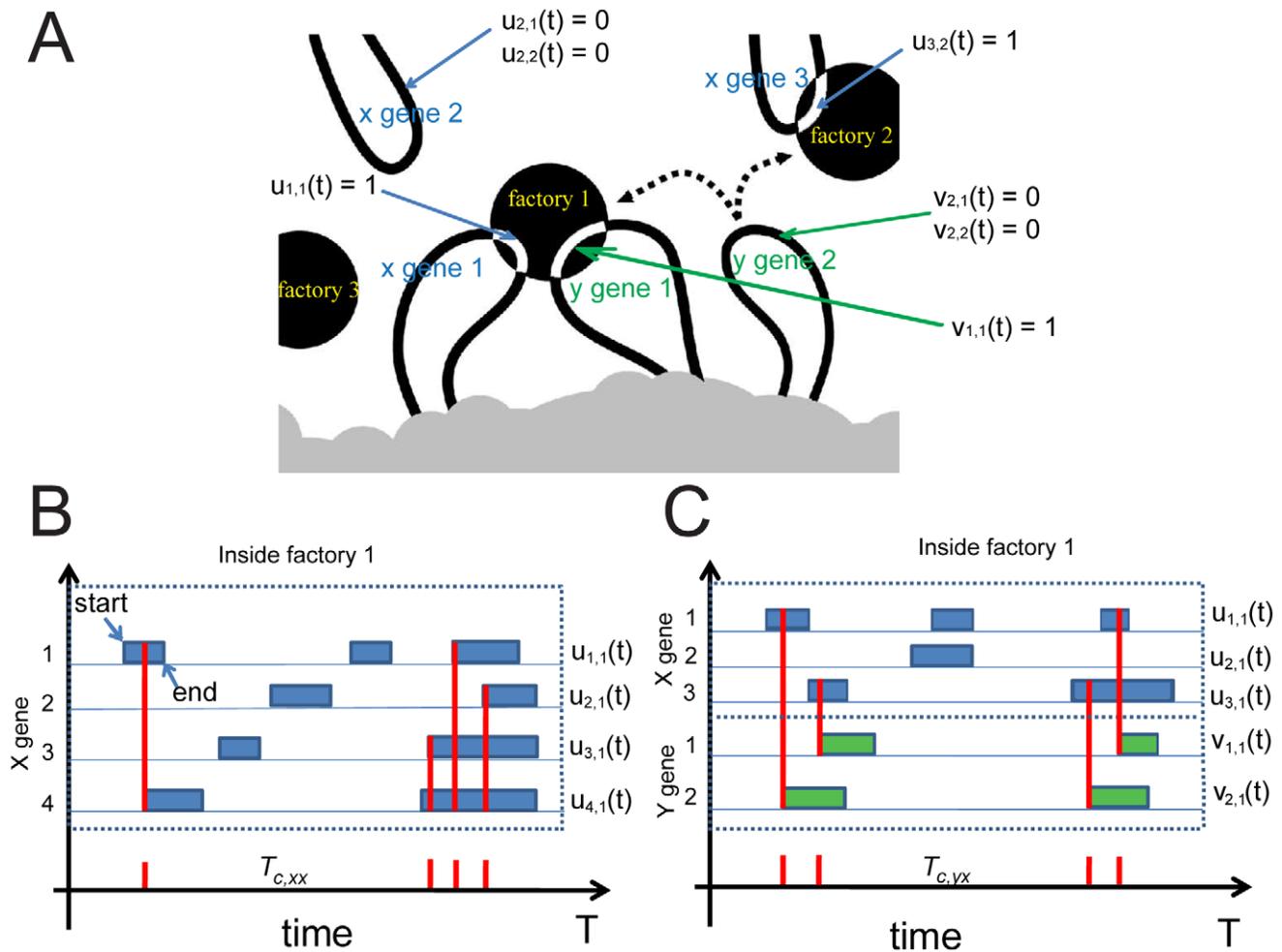

**Figure 1. Demonstration of positionings and states of gene and transcription factory.** (**A**) Schematic representation of chromatin loops (black) extruding from a chromosome territory (gray). Transcribed genes (white) in RNAP II factories (black circles). Potentiated genes (free loops) that are not associated with RNAP II factories are temporarily not transcribed. Potentiated genes can migrate to a limited number of preassembled RNAP II factories to be transcribed (dotted arrows). Both cis and trans associations are possible. If a piece of gene is being transcribed at a particular time $t$, we define the state of that piece of gene as 1 (e.g., $u_{1,1}(t) = 1$, where the first subscript indicates the specific gene, and the second subscript indicates the specific factory that the gene is associated with), otherwise, the state of the gene is defined to be 0. Hence, $u_{1,1}(t) = 1$ means X gene 1 is being transcribed at factory 1, $v_{1,1}(t)$ means Y gene 1 is being transcribed at factory 1, and $u_{3,2}(t) = 1$ means X gene 3 is being transcribed at factory 2. Note that in our simulation we fixed the transcription time for each gene to be 5 minutes. However a gene might be transcribed for longer than 5 minutes if another factor binding event occurs, so that the whole transcription process starts again (i.e. re-initiation). This is illustrated in the figure why genes might have various transcription time. (**B**) Illustration of the co-localization events (red lines) among the same family of genes (X gene) within time window [0 T] (in dashed line window) in factory 1. Once there is a gene start being transcribe inside factory 1 while there are one or more than one genes already being transcribed inside the same factory at time $t$, we say there is a co-localization event happened at time $t$. Therefore, we have 4 co-localization events among X genes in total over time $T$, i.e. $N_{xx}(1,[0,T]) = 4$. The inter-co-localization interval $T_{c,xx}$ tells the timing of X-X gene co-localization. (**C**) Illustration of co-localization events between different families of genes (X gene and Y gene) within time window [0 T] in factory 1. The X-Y gene co-localization events are similarly defined as X-X gene, as shown by the figure, where we have 4 co-localization events between X gene and Y gene over time $T$, i.e. $N_{yx}(1, [0\ T]) = 4$, and $T_{c,yx}$ is the inter-co-localization interval between Y gene and X gene.
doi:10.1371/journal.pcbi.1002094.g001

**Standard Brownian motion: 2D model.** We start with the 2D model using standard Brownian motion to simulate the translocation of genes and factors. All parameters were obtained from experimental data [4,25]. When a gene engages with a factory for a productive transcription event, the presence of other genes transcribing in that factory results in a co-localization. The simulation is run until both X-X gene co-localization events and X-Y gene co-localization events have happened over 5000 times, for each set of parameters. As a result, the co-localization ratio (Eq. (13)) becomes rapidly stable with a small fluctuation over long

**Table 1.** Parameter values used for 2D and 3D Brownian motion and fractional Brownian motion.

| | Brownian motion | | Fractional Brownian motion ($B_H$) | |
|---|---|---|---|---|
| | 2D | 3D | 2D | 3D |
| Gene family | X gene (Hbb, Hba, Hmbs, Epb4.9), Y gene (Cpox) | X gene; Y gene | X gene; Y gene | X gene; Y gene |
| Factor family | X factor, Y factor | X factor, Y factor | X factor, Y factor | X factor, Y factor |
| # of genes for each family | 40 | 100 | 40 | 100 |
| # of factors for each family | 5–12 | 5–25 | 5–12 | 5–25 |
| # of factories | 9 | 35 | 9 | 35 |
| Binding Time $T_b$ (sec) | 10–50 | 10–150 | 10–50 | 10–150 |
| Gene diffusion rate $\sigma_g^2$ | 0.001 ($\mu m^2/s$) | 0.001 ($\mu m^2/s$) | 0.001 ($\mu m^2/s^{2H}$) | 0.001 ($\mu m^2/s^{2H}$) |
| Factor diffusion rate $\sigma_T^2$ | 0.01 ($\mu m^2/s$) | 0.01 ($\mu m^2/s$) | 0.01 ($\mu m^2/s^{2H}$) | 0.01 ($\mu m^2/s^{2H}$) |
| Transcription time $T_t$ (sec) | 300 | 300 | 300 | 300 |
| Gene radius $r_g$ ($\mu m$) | 0.01 | 0.025 | 0.01 | 0.025 |
| Factor radius ($\mu m$) | $r_g/3$ | $r_g/3$ | $r_g/3$ | $r_g/3$ |
| Factory radius ($\mu m$) | $r_g*10/3$ | $r_g*10/3$ | $r_g*10/3$ | $r_g*10/3$ |
| Gene radius after binding with factor ($\mu m$) | $1.5*r_g$ | $1.5*r_g$ | $1.5*r_g$ | $1.5*r_g$ |

doi:10.1371/journal.pcbi.1002094.t001

simulation time (Fig. 2A), as pointed out in the Methods section. The parameter values in Fig. 2A are specified to be 40 X genes, 40 Y genes, 5 X factors, 5 Y factors, and 10 sec binding time. The yellow triangles are the co-localization ratio obtained from Eq. (15), which used the inter-co-localization interval for the calculation rather than the counting of events. It shows that the two methods (Eq. (13) and Eq. (15)) for calculation of the co-localization ratio are equivalent.

In Fig. 2B, the histogram of inter-co-localization interval ($T_{C,xx}$ and $T_{C,xy}$) is plotted for X-X and X-Y gene co-localization, which can be well fitted by gamma distributions. The distribution of X-X gene inter-co-localization interval $T_{C,xx}$ is much narrower than $T_{C,xy}$, showing that X-X gene co-localization events are more likely to happen over time. This means co-localization is more often observed between genes of the same family. Fig. 2C further demonstrates the asymmetry of co-localization for homogeneous and heterogeneous families of genes, by pseudocolor plotting the mean values of inter-co-localization interval $ET_{c,xx}$ and $ET_{c,xy}$ for various parameters (4–11 factors and 5–100 seconds binding). The mean of inter-co-localization interval of genes from the same family ($T_{C,xx}$) are generally smaller compared with that from different gene families ($T_{C,xy}$) for each parameter set, indicating the same family of gene tends to co-localization more often. It is also observed that the inter-co-localization interval increases dramatically as the binding time increases, for both cases, when the number of factors is fixed.

Moreover, Fig. 2D demonstrates the probability of gene-gene co-localization in the 2D case for one specific parameter combination (7 factors, 1 sec binding time). The co-localization probability can be understood as the chance when a gene starts to transcribe that another gene is already being transcribed in the same factory. The histograms of co-localization probability between X-X genes (Eq. (18), pink) and X-Y genes (Eq. (19), blue) show that X-X genes colocalize preferentially, compared to X-Y genes, because the chance for X-X gene co-localization are clearly higher than that of X-Y gene (i.e., $r_2(x_1, x_2|x_1) = 0.033 > r_2(x_1, y_1|x_1) = 0.02$, see Methods section). Finally, Fig. 2E illustrates the X gene's co-localization ratio $r_{xx/xy}$ (Eq. (13)) for various parameters, calculated from counting of co-localization events for X-X gene and X-Y gene. The red dashed line indicates the classification threshold for co-localization region where $r_{xx/xy} > 0.52$ (region above). Interestingly, although the model is based on completely random movements of genes and factors, there is a region of parameters in which the system behaves non-randomly, where X genes tend to colocalize preferentially with other X genes.

The interesting question is where the asymmetry (i.e., tendency for genes of the same family to colocalize) comes from. Consider an extreme case in which the number of transcription factors for each family is severely limited (i.e., one X factor and one Y factor). In this case, only family member genes in the immediate vicinity of their factor have a chance of transcribing and co-localization of the family member genes in the same factory would have a high probability. On the other hand, if there were enough factors to bind every gene simultaneously, gene transcription events would be totally independent and a random distribution of factory sharing would be expected. Therefore, for a fixed number of factories, the co-localization ratio is a decreasing function of the number of transcription factors, as indicated in Fig. 2E, while gene-factor binding time does not influence gene co-localization much. Hence, limited resources (finite number of transcription factors) could be the main reason for co-localization in the simulation.

**Standard Brownian motion: 3D Model.** Little information is available about how gene regulatory components are organized within the three-dimensional space of the nucleus from experiments. Our 3D model can provide a temporal-spatio simulation of the translocation of genes and transcription factors, and their interactions with factories. Video demonstration for genes and factor translocation, binding process and transcription in the 3D case (when the locations of transcription factories are fixed) is available from our group's website (www.dcs.warwick.ac.uk/~feng/gene-factory.avi). Fig. 3A illustrates the positioning of each transcription element at a specific time $t$ (big green dots represent the location of each factory, asterisks are the transcription factors, and genes the small spots. Red and blue represent different family of genes and factors). The parameters of

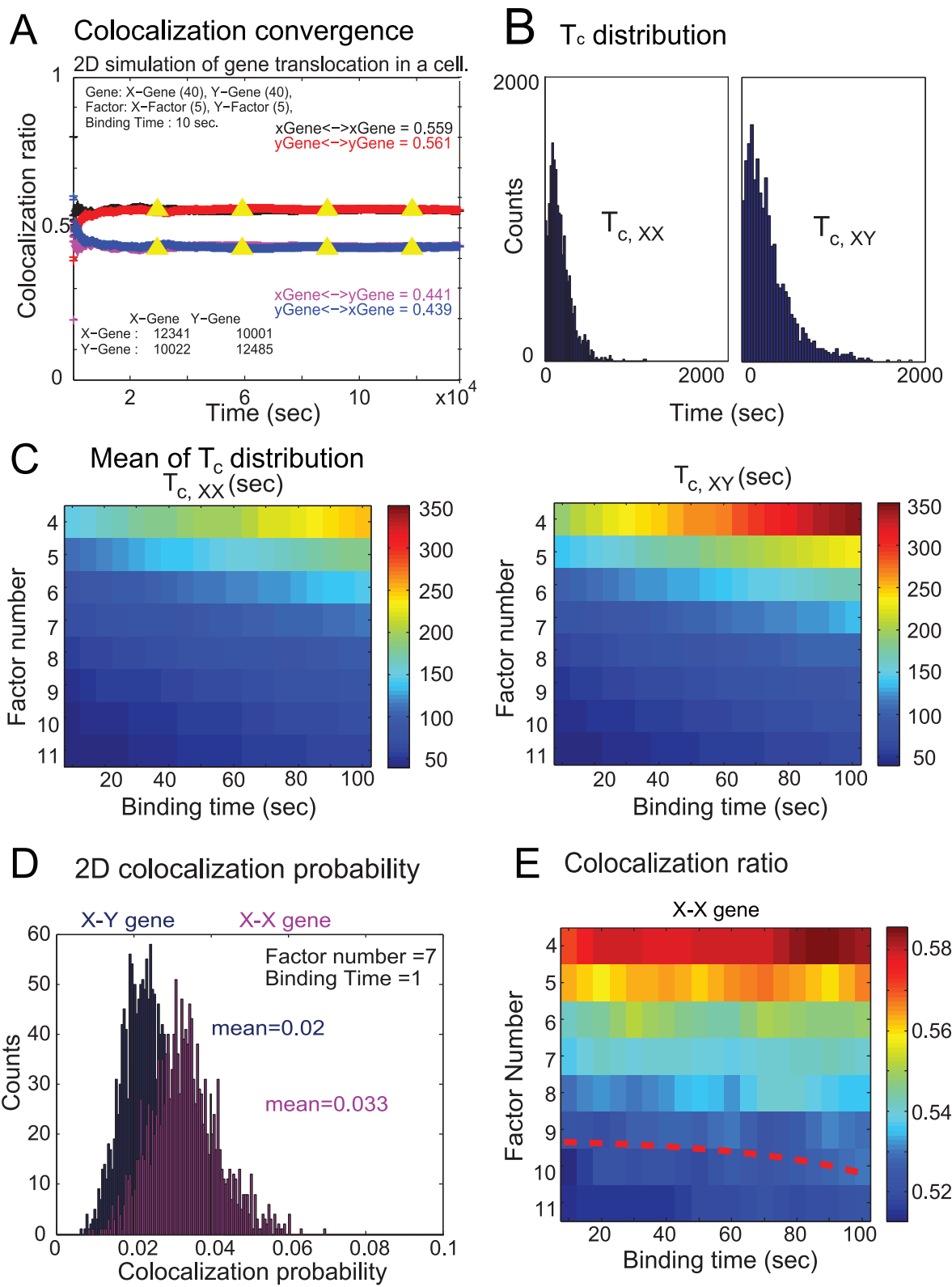

**Figure 2. Simulation results for 2D case.** (**A**) One example of co-localization ratio convergence. The ratio rapidly becomes stable as time increases. Yellow triangle is obtained from Eq. (15), showing the consistency of calculation of co-localization ratio using different variables. (**B**) The inter-co-localization (ICI) interval distribution for $T_{c,xx}$ and $T_{c,xy}$. The parameters are the same as displayed in (A). The inter-co-localization interval distribution can be fitted with gamma distributions. (**C**) The mean values of ICI distribution by varying the factor number and binding time between genes and factors. The mean ICI increases as the binding time increases, and decreases as the factor number increases. (**D**) The co-localization probability for X-X gene and X-Y gene. (**E**) Co-localization ratio for various combinations of factor number and binding time. The ratio threshold is set to be 0.52. The red dashed curve is the threshold boundary to distinguish whether the co-localization is significant or tends to be random.
doi:10.1371/journal.pcbi.1002094.g002

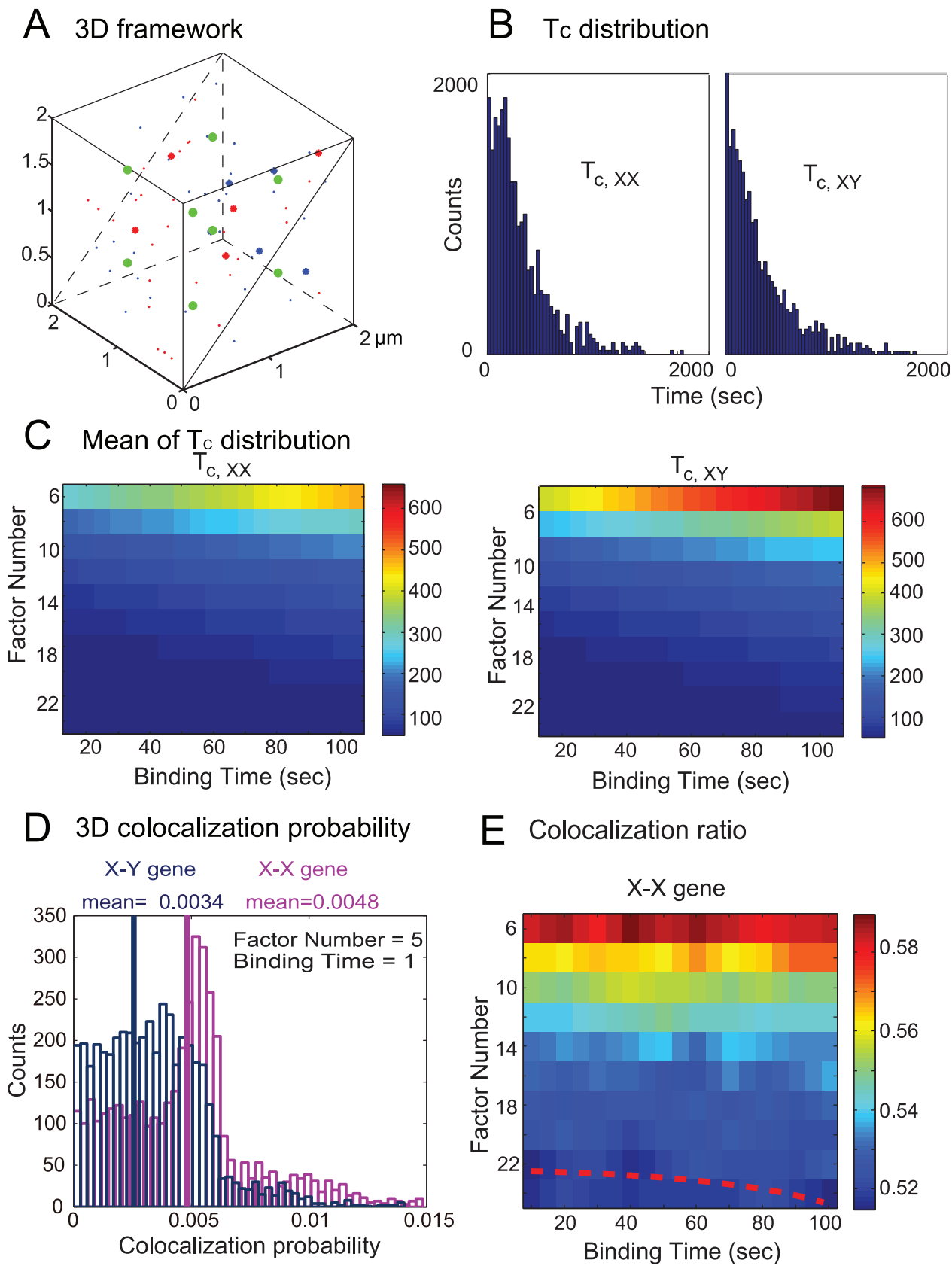

**Figure 3. Simulation results for 3D case, similar as Fig. 2.** (**A**) Demonstration of the 3D framework in a cubic, where the big green dots represent the location of the transcription factories, asteroids the transcription factors, and small dots the genes. Different families are represented by either red or blue. (**B**) The inter-co-localization interval distribution for $T_{c,xx}$ and $T_{c,xy}$. The parameters are the same as in (D). (**C**) The mean values of ICI

distributions by varying the factor number and binding time between genes and factors. The mean ICI increases as the binding time increases, and decreases as the factor number increases. (**D**) The distribution of the co-localization ratio for X-X genes and X-Y genes. (**E**) The colicalization ratio for various combinations of factor numbers and binding times. The critical ratio is 0.52. The red dashed curve is the threshold boundary to distinguish whether the co-localization is significant or tends to be random.
doi:10.1371/journal.pcbi.1002094.g003

3D simulation can be found in Table 1. Most of the parameters are the same as the 2D case, except the genes and factors translocate randomly in a cubic space rather than a square plane (Fig. 3A).

Fig. 3B–E are obtained similarly as Fig. 2B–E. Interestingly, in comparison with the 2D case, the area of the co-localization ratio is considerably enlarged (Fig. 3E). This could be easily understood from the dynamics of the 3D Brownian motion. The main reason for the bias (co-localization) is again due to the limited number of transcription factors, as discussed in the 2D case previously. With Brownian motion in the 3D cube, it is more difficult for genes and factors to collide and combine with each other, and then engage with a factory compared to the 2D square. Hence the probability that transcription factors of two different families cluster in the same factory is lower. This explains why the co-localization regions in Fig. 3E are larger than Fig. 2E.

**Fractional Brownian motion.** Is standard Brownian motion good enough to simulate preferential co-localization in nucleus? Experimental data of chromatin movements reported in yeast cells in Sage et al [15], Gasser [26] and Heun et al [27] demonstrates that fractional rather than standard Brownian motion may be more appropriate to model gene mobilizations. In S and G1 phase of the cell cycle, the trajectory of the chromatin movement is negatively correlated (one long increment followed by a short one), and therefore fractional Brownian motion is a sub-diffusion process (see Methods section). This means the trajectory of the chromatin movement is more localized, enabling the genes to colocalize more easily.

With the information from experiments, we expect that fractional Brownian motions play an important role in the nuclear dynamics. To test the hypothesis, we ran simulations in 2D and 3D using fractional Brownian motion to simulate the translocation of genes and factors. Fig. 4A is the inter-co-localization interval histogram for the sub- and super-diffusion with 5 factors and 10 seconds binding time in 2D. Clearly, the inter-co-localization interval of sub-diffusion ($H=0.1$) is much smaller than super-diffusion ($H=0.9$), indicating that co-localization between members of the same family of gene occurs more readily for sub-diffusion ($H=0.1$, as $T_{C,xx} < T_{C,xy}$ for both cases). Fig. 4B shows the co-localization region for sub-diffusion ($H=0.1$) and super-diffusion ($H=0.9$) movement in 2D. It clearly reveals that the co-localization region is very large for sub-diffusion as the co-localization ratio is always above the threshold 0.52, but for super-diffusion, co-localization is hardly observed. Furthermore, Fig. 4C illustrates that the co-localization ratio is a monotonically deceasing function of $H$ in 3D. Intuitively, the average target hitting time for sub-diffusion movement should be the same as super-diffusion, but for genes moving with sub-diffusion, the chance of a gene to re-enter the same factory after exiting that factory would be much larger than with super-diffusion, since the particle will stay locally and hit the target once again more easily. In comparison, gene co-localization level under super-diffusion process would not be affected much as genes tend to move more globally. This may suggest that sub-diffusion in gene translocation is biologically significant (as indicated from the data obtained from [15]).

## Co-localization (association level) between genes and factors

All results above tell us that there exist co-localization regions between genes, even though the model is set up completely symmetric (equal number of genes and factors for each family). We further compared our simulation results with recently obtained experimental data (Fig. 5A) to validate our model.

Schoenfelder et al [7] reported the intra- and inter-chromosomal co-localization frequencies of 33 mouse genes relative to the Hbb and Hba globin genes in erythroid tissues (Fig. 5A). Gene regulated by the transcription factory Klf1 preferentially clustered in factories containing high levels of Klf1. Fig. 5Aa shows the spatial distribution of transcription factor Klf1 (Kruppel like factor) relative to RNAPII factories by immunofluorescence in mouse erythroid cells. The data exhibits nearly all nuclear Klf1 foci overlapped with RNAPII-S5P foci, indicating 10–20% of transcription factories contain high levels of Klf1. Therefore, we restrict 20% of Klf1 associated level with RNAPII (as the background association level) by selecting the factor number and binding time in our simulation (Fig. 5B). Fig. 5Ab is the double-label RNA immune-FISH of nascent transcripts (Hbb, green) and Klf1 foci (red). This image shows the positions of transcriptionally active, Klf1-regulated gene (e.g., Hbb) relative to Klf1 foci. It is found that majority (59%–72%) of actively transcribed alleles of Hbb, Hba, Hmbs and Epb4.9 (regard as X genes) were preferentially associated with Klf1 transcription factories. Cpox genes (regard as Y gene) associate with Klf1 factories at marginally higher frequencies (26%) than expected by a purely random distribution. For actively transcribed alleles of the Klf1-independent Tubb5 and Hist1 genes (regard as Z genes), they show no preferential localization to Klf1-containing factories (20%). Hence, we regard Klf1 as the X factor, and the X gene - X factor association level is estimated to be around 64% in experiments, while Y gene – X factor association level is around 20% from this experiment, matching the Klf1 background association level (20%). Therefore, we understand that X factors and Y genes are independent to each other. Fig. 5Ac is the triple-label RNA immune-FISH for pairs of nascent transcripts (Hbb and Hist1, blue and green, respectively) and Klf1 foci (red). From experimental observation, this colocalizing pair of genes relative to Klf1 foci reveal that colocalizing pairs of Klf1-regulated genes are associated with Klf1 transcription factories at very high frequencies (63–79%), and the colocalizing Klf1-independent gene pairs show no preferential association with Klf1 transcription factories.

In simulation, we calculated the association level between X gene (Hbb, Hba, Hmbs, and Epb4.9) and X transcription factor (Klf1), the association level between Y gene (Cpox) and X factor (Klf1), when confining the Klf1 background association level as 20%. Using sub-diffusion ($H=0.4$) to simulate the translocation of regulatory elements, we calculated the X gene – X factor association level (Eq. (25)) both in 2D and 3D, by fixing gene number but varying factor number and binding time (Fig. 5Ba), or by fixing factor number but varying gene number and binding time (Fig. 5Bb). It is clearly shown that the experimental data can be well matched with our model with one set of parameters (gene number 5, factor number 2, and binding time 130 s) in 3D case.

Next we examine the co-localization between a pair of genes (X genes, Y genes or Z genes) and Klf1 (X factor). Our simulation result shows that the co-localization of paired X genes with Klf1 is 0.8, and paired X-Y genes with Klf1 is 0.6, again in agreement with experiments [7]. All experimental results and our simulation results except for Z genes are summarized in Table 2.

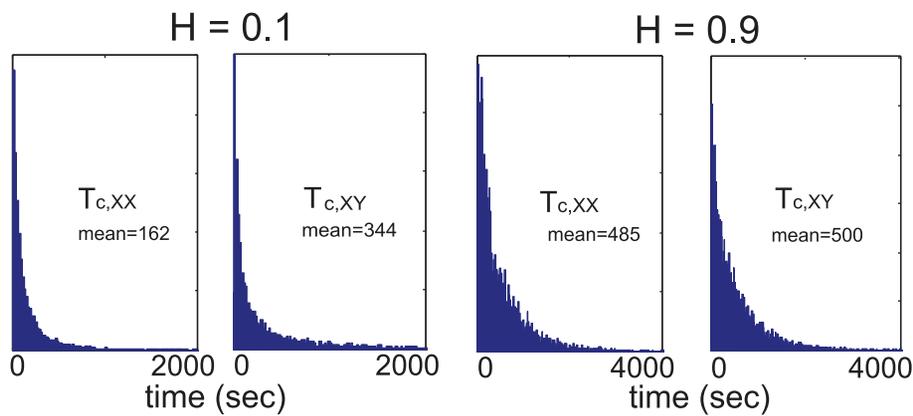
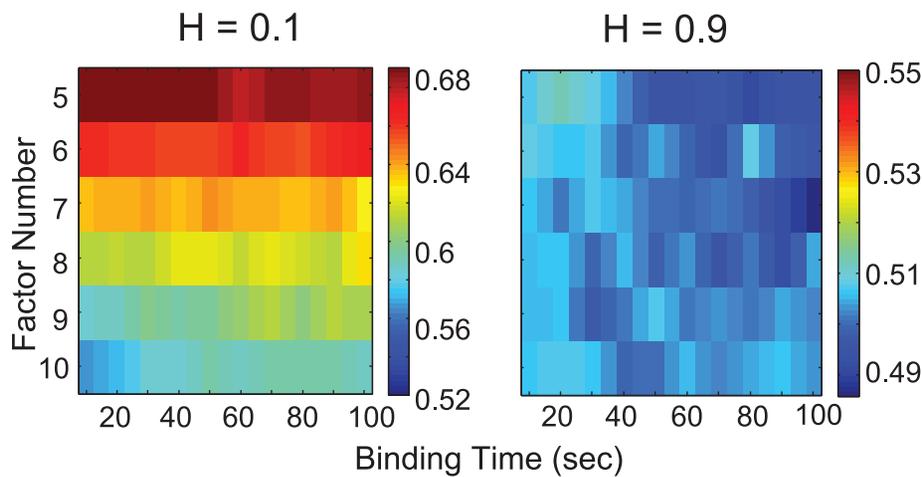
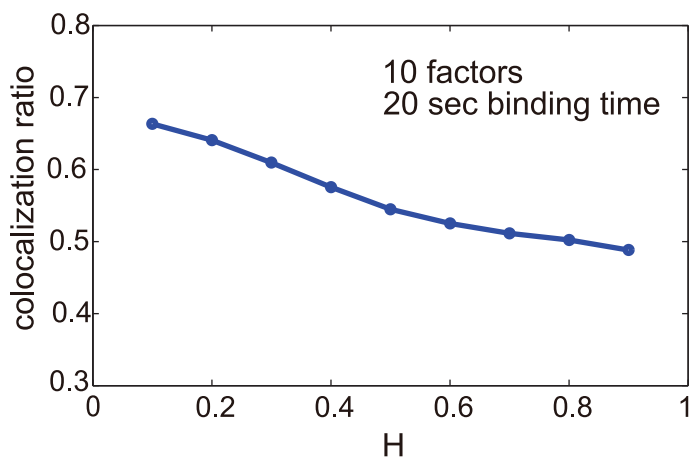

**Figure 4. Co-localization regions for sub- and supperdiffusion Brownian motions.** (A) The inter-co-localization intervals for super- and sub-diffusion for 5 transcription factors and 10 sec binding time. (B) Co-localization regions for different factors and binding times for sub-diffusion case ($H=0.1$) and super-diffusion case ($H=0.9$) in 2D. (C) Co-localization ratio versus the Hurst index $H$ in 3D.
doi:10.1371/journal.pcbi.1002094.g004

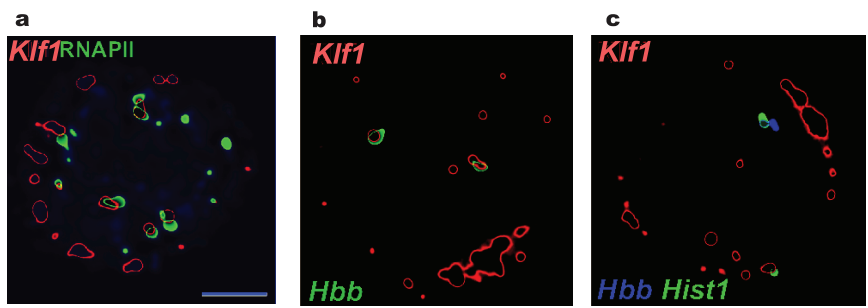

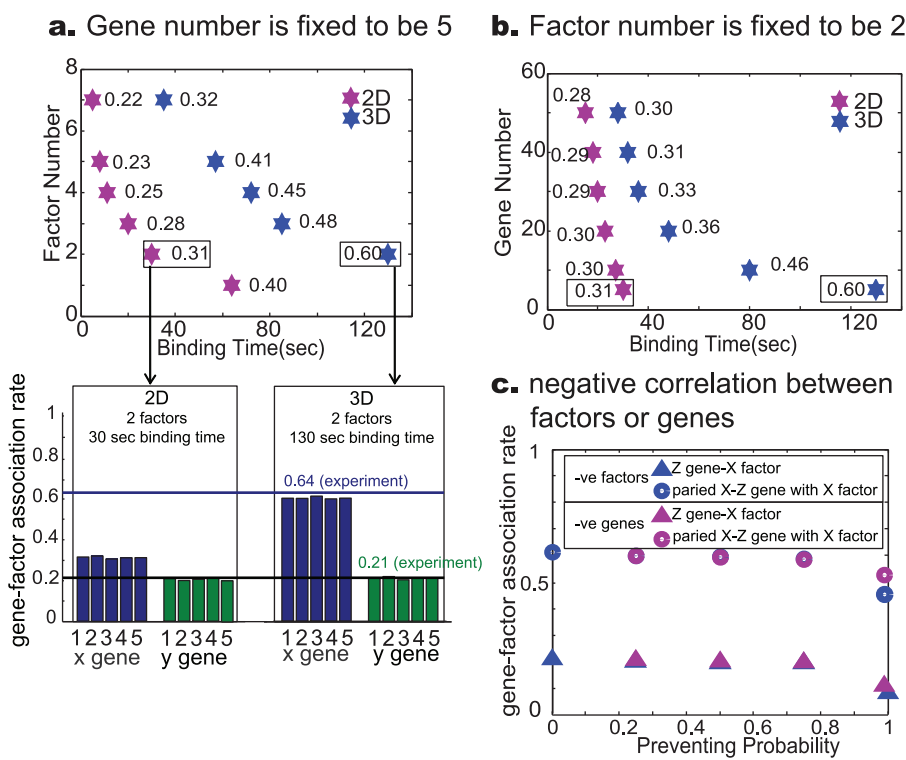

Figure 5. Co-localization between genes and transcription factors. (A) Experimental results with immuno-RNA FISH reveal co-localization between transcription factors and factories (RNAPII), genes and transcription factors, and gene pairs and transcription factors. This is the contour plot from original experimental data. (Aa) Immune-fluorescence detection of Klf1 (red) and RNAPII-S5P (green) in definitive erythroid cells, with a scale bar of 2 μm. This shows the co-localization between Klf1 and transcription factories RNAPII. This Klf1 background association rate (level) is estimated to be 20%. (Ab) The co-localization between transcription factor Klf1 and Hbb gene. (Ac) The co-localization between factor Klf1 and genes pairs (Hbb and Hist1). (B) Simulation results when we hold the Klf1 background association level as 20%, while the translocation of transcription units (genes, factors and factories) are following sub-diffusion process ($H = 0.4$). (Ba) Gene-factor association level (numbers indicated behind the stars) with various factors and binding time, both for 2D and 3D cases. The number of genes (for each family) is fixed to be 5 and the Klf1 (X factor) background association level is fixed to be 20% (the stars indicate the parameter values when this condition is satisfied). The detailed Klf1 association level for each X gene and Y gene are presented in the figure below, revealing the fact that the simulation results for 3D case (gene-factor association level = 0.6) match the experimental result (gene-factor association level = 0.64) quite well for a specific set of parameters (5 genes, 2 factors and 130 sec binding time). (Bb) Gene-factor association level (numbers besides the stars) with various genes and binding time for 2D and 3D cases. The number of factors is fixed to be 2 and the Klf1 background association level is fixed to be 20%. (Bc) The association rate of Klf1 (X factor) with Z gene and paired X-Z genes when there is a negative correlation between X and Z gene, or X and Z factors, under 3D case. The parameter used here are 5 genes, 2 factors, and 130 sec binding time. Preventing probability means the chance for stopping another gene (factor) to enter the factory when there is already one gene (factor) in that factory. When $p = 0$, it represents the independent situation of X factor (gene) and Y factor (gene).
doi:10.1371/journal.pcbi.1002094.g005

**Table 2.** Comparison of sub-diffusion 2D simulation results (Fig. 5b, 3D case) and the experimental results.

|  | Factor background level | Gene-factor association level | | | Paired genes-factor association level | | |
| --- | --- | --- | --- | --- | --- | --- | --- |
|  | Klf1 | X | Y | Z | X-X | X-Y | X-Z |
| Simulation | 0.2 | 0.6 | 0.2 | $r2(z1,u^T|z1)$ | 0.8 | 0.6 | $r3(u1,z1,u^T|u1,z1)$ |
| Experiment | 0.2 | 0.64 | 0.2 | 0.2 | 0.8 | 0.64 | 0.2 |

doi:10.1371/journal.pcbi.1002094.t002

We tried to understand why X-Z gene pairs (Hbb/Tubb5, Hba/Tubb5, Hbb/Hist1, Hba/Hist1) association level with Klf1 factors is low. One method is to introduce interactions (negative correlation) between X gene transcription factors and Z gene transcription factors or X genes and Z genes themselves. In other words, Z genes (or Z factors) might be negatively correlated with X genes (or X factors), while Y genes and X genes are independent. To assess this, we ran simulations with the following exclusive rules: if an X gene (factor) is in a factory, it will prevent the entry of a Z gene (factor) with a probability $p$. In factor case, it simply implies that Z gene is co-regulated by X and Z factors. The simulation results on the gene-factor association level (Fig. 5Bc) did not show much difference after including the preventing probability among genes or factors, and it is not easy to simultaneously fit the experimental data which implies that Z gene – X factor association level as 0.2, indicating the negative correlation between different families of genes (factors) might not be the primary reason for different values of Klf1 association rate among different families of genes, as observed from experiments [7]. Hence, more sophisticated interactions are required, and we will further investigate this phenomenon in our future work.

### Nucleus size, factory numbers and factor numbers

Osborne et al. [1] shows approximately 2000 transcription factories in the extended and flattened nuclei of mouse embryo fibroblast. In contrast (Fig. 6A), they found that erythroblast, B cell, T cell and fetal brain cells, which have spherical nuclei with significantly smaller radii and nuclear volumes, have dramatically fewer transcription factories (100–300 per nucleus). It was argued that the large differences in factory numbers seen in nuclei from tissues versus cells grown on a surface appear genuine and may be a consequence of a reduced potential for inter-chromosomal sharing of factories in flattened cell nuclei [28]. To test how the changes in nuclei shape and transcription factory number will effect on gene co-localization, we ran simulations with flattened cells, squashing the original cubic from 2×2×2 to 0.5×4×4 but maintaining its volume (Fig. 6A). We have also tested the situation when the flattened cell is of volume five times bigger than the spherical cell $(0.5 \times \sqrt{80} \times \sqrt{80})$, according to the experimental observation (data available by request). We partitioned the flattened cell into four subunits (0.5×2×2) and restricted the translocation of genes within each subunit, so all genes are restricted locally for consistency with experiments while transcription factors are free to move within the entire space. We found, consistent with our 'limited resource' theory, that no matter if we increase the volume of the flattened cell or not, the colocalized transcription is increased rather than reduced in the flatten cell, and is almost independent of the number of transcription factories (Fig. 6A). Moreover, it is also observed from Fig. 6A that increasing the volume of nucleus might increase the co-localization ratio. We will investigate the reason mathematically in our future research work.

From the analysis above, we propose the possibility to reconcile the facts observed in experiments and our model simulations: increasing the transcription factors might be the only possible mechanism to prevent gene co-localization. To confirm this, we ran the simulations with different number of transcription factors (Fig. 6B), for different volumes of the flattened cell. The left panel of Fig. 6B (unchanged volume) clearly demonstrated that when the number of transcription factors is around 30, the colalization ratio is reduced to around 0.5 (non-co-localization) and is independent of the number of transcription factories. When the nuclear volume is enlarged 5 times bigger that of the original spherical nucleus (Fig. 5B right panel), co-localization is even easier to happen for various cases, but increasing the number of transcription factory can hardly be the only reason for higher chance of co-localization.

### Frequency- and amplitude-modulation in nucleus

In eukaryotic cells, external signals can modulate the expression of target genes by regulating the nuclear versus cytoplasmic localization of transcription factors. Experimentally, we have observed two possible types of modulations: one is amplitude modulation, implying that external signals regulate a static number of transcription factors into the nucleus (Klf1 might be an example [7]); the other is frequency modulation, in which external signals alter the frequency of nuclear bursts (entry/exit cycles) of the transcription factor (for example, p53 [29], NF-κB [16] and Crz1 [17]).

Cai et al observed that the nuclear localization burst frequency of Crz1, a transcription factor that regulates more than 100 target genes, increases in response to the increase in extracellular calcium concentration [17]. In addition, they suggested and experimentally verified that this frequency-modulation mechanism of transcription factor localizationcan coordinate the expression of multiple target genes, whereas amplitude-modulation cannot.

We assessed whether co-localization is affected by these two different modulations, using our model. Our previous model setting is equivalent to a (fixed) amplitude modulation scenario where the number of transcription factors is kept as a constant in the nucleus. We investigated whether frequency-modulation of factors could be involved in the control of multiple target gene co-expression compared to amplitude-modulation. In our simulations, we regard Crz1 as X factor (no Y factor is present), and assume that Crz1 binds to two families of target genes ($X_1$ gene and $X_2$ gene), which have completely different diffusion rates, radiuses and transcription times (binding time as 10 sec, and other detailed parameters are presented in Table 3). We ran our simulations in 3D cubic with a sub-diffusion ($H = 0.4$).

Fig. 7A demonstrates the dynamics of transcription factor translocation into and out off the nucleus with various frequencies, and Fig. 7B shows the factor entry profile into nucleus under frequency-modulation. Each period is composed of an active burst part and refractory part. During the refractory time, only very few (residual) factors are in nucleus, and as a result, only very few genes can be transcribed. While in the active burst time (2 min as reported in [17]), many transcription factors swarm into the

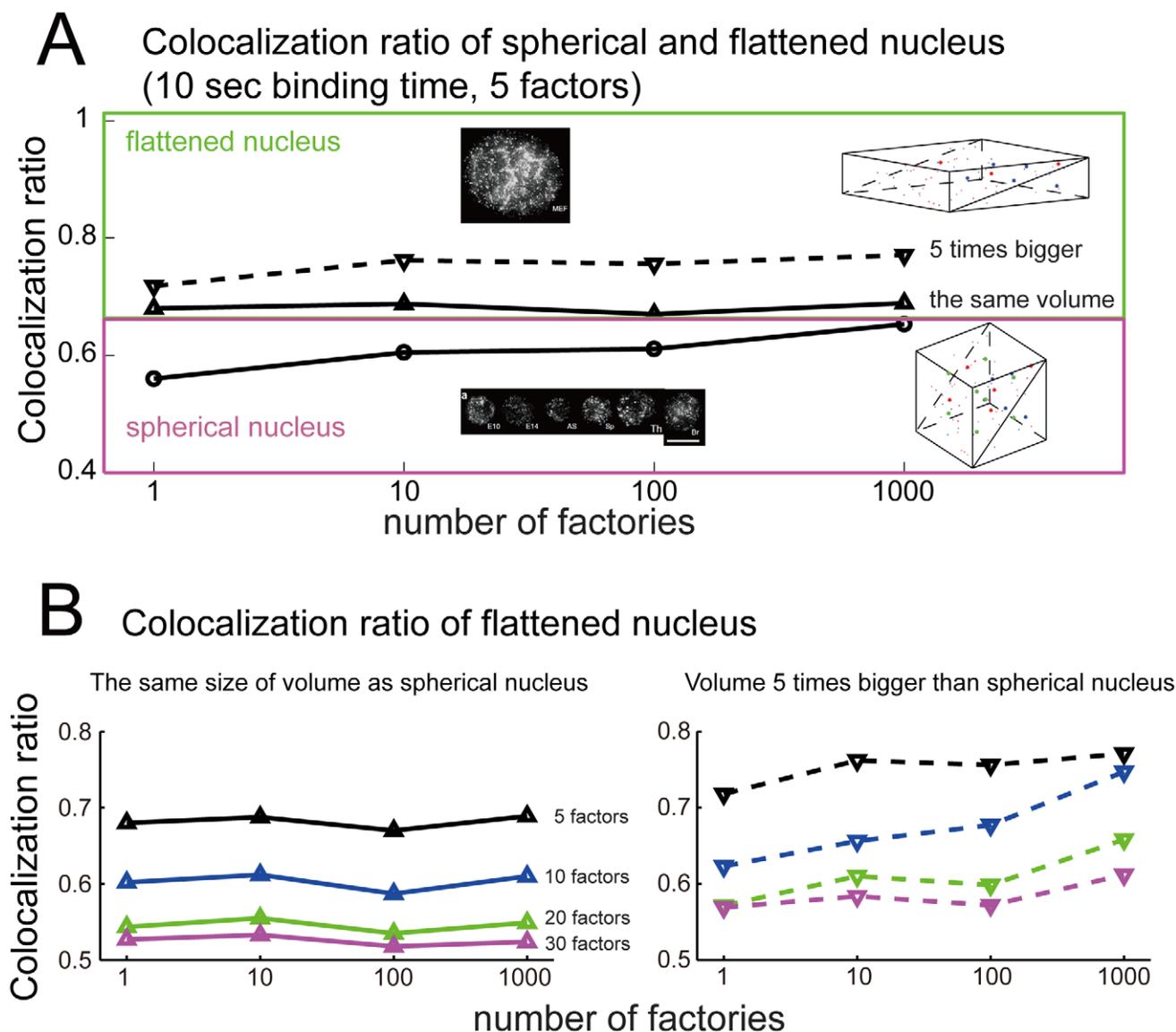

**Figure 6. The simulation results for ratio of co-localization in flattened and spherical nucleus.** (**A**) The co-localization ratio increases as the degree of flatness of the nucleus increases when there are 5 factors and 10 sec binding time, and is independent of the number of transcription factories, at least in the flattened nucleus, such as E10 (embryonic blood), E14 (fetal liver erythroid), AS (adult anemic spleen erythroid), Sp (normal adult spleen), Th (adult thymus) , Br (fetal brain), mouse embryonic fibroblasts (MEFs) in experiments. Scale bar = 10 µm. The cubic of spherical nucleus and rectangular block of flatten nucleus demonstrate the positioning of transcription factories, factors and genes (refer to Fig. 3A). Note that the volumes of the cubic spherical nucleus and the rectangular flattened block are either the same (solid line), or the volume of the flattened nucleus is 5 times bigger than that of the spherical nucleus (dash line). (**B**) The co-localization ratio is a decreasing function of the number of transcription factors for both flattened nucleus of the same volume as the spherical one, and the nucleus of 5 times larger volume. No matter the volume, the co-localization ratio is independent of the transcription factories.
doi:10.1371/journal.pcbi.1002094.g006

**Table 3.** The parameters for two different genes under amplitude modulation and frequency modulation.

| H = 0.4 | $X_1$ gene | $X_2$ gene |
|---|---|---|
| Gene diffusion rate $\sigma_g^2$ (µm$^2$/s$^{2H}$) | 0.0005 | 0.0015 |
| Transcription time $T_t$ (sec) | 50 | 150 |
| Gene radius $r_g$ (µm) | 0.05 | 0.016 |

doi:10.1371/journal.pcbi.1002094.t003

nucleus, and diffuse as the model described in the amplitude modulation case.

In Fig. 7C and Fig. 7D, we illustrate the evolution of the normalized expression level of two kinds of genes ($X_1$ and $X_2$ gene) as a function of the factor amplitude and burst frequency, respectively. Clearly, $X_1$ gene and $X_2$ gene yield uncoordinated expression patterns under amplitude modulation; while the curves of $X_1$ and $X_2$ gene normalized expression levels almost coincide under frequency modulation, as suggested in [17].

Now we are in the position to assess the impact of frequency and amplitude modulation on gene co-localization. To this end, we have two types of factors and corresponding genes. One type is frequency modulated (X gene and X factor), the other is amplitude modulated (Y gene and Y factor). Like the symmetric model we simulated in above sections, X factor and Y factor bind to X gene and Y gene respectively. Besides, X gene and Y gene have identical properties (parameters are in Table 1). For comparability, the average number of X factors in the whole simulation time should be the same as the static number of Y factor. Unlike the previous symmetric models, here X gene co-localization ratio differs greatly from Y gene co-localization ratio, as indicated in Fig. 7E. For all burst frequencies, X gene's co-localization ratio is larger than Y gene, which implies that genes regulated by frequency modulated factors may colocalize in the nucleus more than genes of amplitude regulated factor. Although in our simulation there is almost no X-X gene co-localization event in refractory time of burst, X-X gene co-localization event in active time happens more than Y-Y genes since the average number of X factor in active time is larger than Y factor number. We conclude that in additional to the coordination of target gene expressions, another functional role of frequency modulation of factor entry may be to facilitate co-localization between target genes. This can be one interesting biological experiment to evidence whether frequency modulation allows higher co-localization and higher levels of coordinate expression of groups of genes.

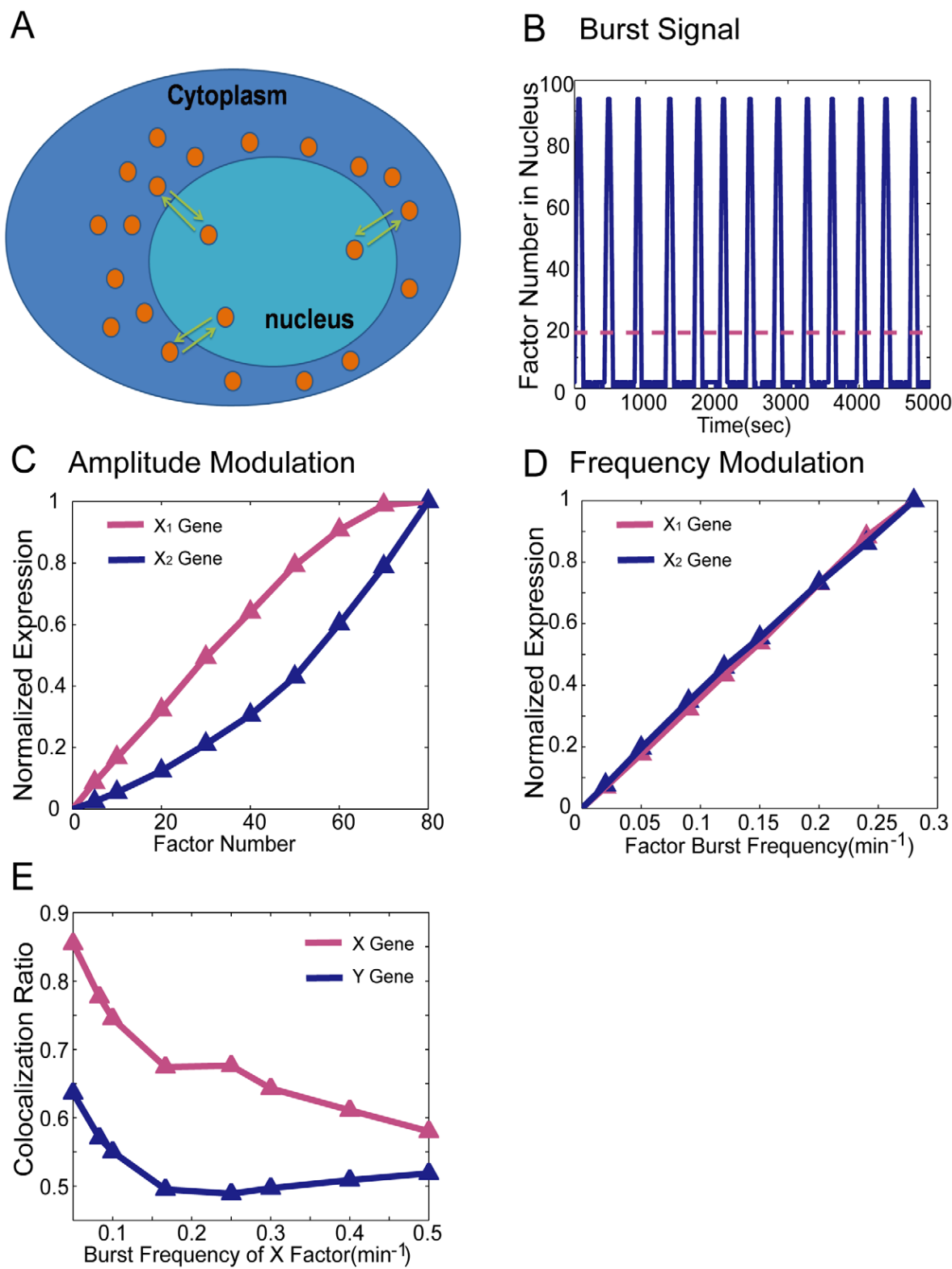

**Figure 7. Amplitude modulation and frequency modulation of factor number.** (**A**) Illustration of transcription factor translocation into and out of the nucleus. (**B**) The burst of factor nuclear localization under the frequency of 0.15(min$^{-1}$) The red dash line represents the average factor number over each period. (**C**) The normalized expression level of $X_1$ gene and $X_2$ gene versus factor number (amplitude modulation). (**D**) The normalized expression level of $X_1$ gene and $X_2$ gene versus burst frequency of factor (frequency modulation). (**E**) X gene and Y gene co-localization ratio versus burst frequency of X transcription factor.
doi:10.1371/journal.pcbi.1002094.g007

## Discussion

In the current paper, we have investigated whether a simple diffusion model can account for the co-localization observed in experiments, based upon parameters measured from experiments. We first assess the ratio of gene co-localization. It is found that the co-localization ratio is determined by the inter-co-localization intervals and is biased. We then applied the theory and numerical simulations to two and three dimensional cases with standard and fractional Brownian motion. We have shown that the experimentally observed co-localization is possible in both two and three dimensional cases and conclude that our dynamical model can match many experimental data.

However, a direct comparison with experimental data is still not easy since we do not have data of the dynamics of multi-genes. All experimental results are static results [7]. With the development of new experimental techniques, we expect that the dynamic data should be available soon. Such data would be valuable for us to understand the interactions between genes.

It is clear that our model is a simplified version of gene mobilization in the nucleus: each gene is treated as independent (fractional) Brownian motion which is only true in local loci and small time intervals (Fig. 1A) [30,31]. The transcription process is also a simplified process. Moreover, we tried to introduce negative correlation between different families of genes and factors with preventing probability $p$. However, the simulation results on the gene-factor association level (Fig. 5Bc) did not show much difference after including the preventing probability among genes or factors. Hence, more sophisticated interactions are required, and we will further investigate this phenomenon in our future work.

In the models above, all genes are treated as a point (point model). Modelling of genes as segments on 3D chromosomes as polymer chains [32] would be more appropriate. The 3D whole genome conformation will be based on Hi-C data (see Lieberman-Aiden et al [33]). The dynamics of each polymer chain will be modeled according to the well known polymer physics [34], in collaboration with our experimental data. The simulation would be computationally very expensive and would therefore need to be run on state-of-the-art clusters. The interactions in the model between genes (chromosomes) etc. should fit well with the known experimental data accumulated in our experimental teams for the past years.

After having a biophysically realistic model (with some coarse-grain approaches), we would expect to use the model to predict some key stages of hematopoietic differentiation. These predictions will then be tested by our experimental groups. Certainly this would be a very challenging task and it is a multi-scale spatio-temporal dynamics. Ideally we should be able to predict key decision making mechanisms at the molecular and cellular level that control genome function and may lead to the lymphoid versus myeloid differentiation. The transcription factories story might fit well with some general computational principle as reviewed in Oehler et al [35].

In general, we have to take into account the interactions between genes, both *in cis* and *in trans*, between genes and transcriptions, and between genes and transcription factories. As mentioned in Methods section (Eq. (1)), we can include the interactions between transcription units in the drift terms [36]:

$$m\frac{d^2 x_i^T(t)}{dt^2} = -\varsigma^T \int_{-\infty}^t \frac{dx_i^T(t)}{dt} K_H(t-s)ds - m\Psi \sum_{j=1}^{N_f}(x_i^T(t)-c_j) + \sqrt{2\varsigma^T k_B T_e}\frac{dB_H(t)}{dt}$$

where $m$ is the mass, $x_i^T$ is the position of transcription factor of X gene, $\Psi$ is the shape parameter of the harmonic potential, $c_i$ is the centre of each factory and $N_f$ is the number of transcription factories, $k_B$ is the Boltzman constant, $\varsigma^T$ is a friction constant, $T_e$ is the temperature and $K_H(t)$ is a kernel so that the fluctuation-dissipation theorem holds true. For each gene, it obeys similar equation

$$m\frac{d^2 x_i(t)}{dt^2} = -\varsigma^T \int_{-\infty}^t \frac{dx_i(t)}{dt} K_H(t-s)ds - m\Psi(x^T(t))\sum_{j=1}^{N_f}(x_i(t)-c_j) + \sqrt{2\varsigma^T k_B T_e}\frac{dB_H(t)}{dt}$$

but with a potential depending on whether its corresponding transcription factor is in a factory or not (the term $\Psi(x^T(t))$). How to find the right parameters of the interactions in the equations above would be an interesting issue. In the past decades, many techniques have been developed, mainly using the idea of Markov chain Monto Carlo and Bayesian approaches (for example, Pavliotis and Stuart [37]). With the drift term introduced here, we could expect that sub-diffusion has a larger co-localization region than super-diffusion.

## Methods

### Standard Brownian motion

Assume that we have $m$ X genes and $k$ Y genes, with $n$ transcription factors of X gene and $l$ transcription factors of Y gene. Denote their positions at time $t$ as

$$\text{X genes}: \quad x_1(t),...,x_m(t);$$

$$\text{X factors}: \quad x_1^T(t),...,x_n^T(t);$$

$$\text{Y genes}: \quad y_1(t),...,y_k(t);$$

$$\text{Y factors}: \quad y_1^T(t),...,y_l^T(t).$$

All genes and transcription factors move according to diffusion processes, i.e.

$$dz(t) = b(\Theta)dt + \sigma_z dB^z(t) \quad (1)$$

where $b$ is the drift term depending on the global activity in the nucleus $\Theta$, $\sigma_z^2$ is the diffusion coefficient of transcription element $z$ (where $z$ can be $x$, $y$, $x^T$ or $y^T$) and $B^z(t)$ is the independent fractional Brownian motion. The genes and transcription factors move around with a constant diffusion coefficient

$$\text{Gene}: \quad \sigma_x^2 = \sigma_y^2 = \sigma_g^2; \quad (2)$$

$$\text{Factor}: \quad \sigma_{x^T}^2 = \sigma_{y^T}^2 = \sigma_T^2. \quad (3)$$

The drift term summarizes the interactions between association of centromeres, clustering of co-regulated genes, association of a regulatory element and its target genes, interaction of a genome region with the nuclear envelope etc [13]. For a gene $h$, define a sequence of stopping (binding) times for X genes and Y genes as

$$\text{X gene}: \quad \tau_h^j = \inf\{t: d(x_h, X^T) \leq \varepsilon_0, t > \tau_h^{j-1} + T_b\} \quad (4)$$

$$\text{Y gene}: \quad \omega_h^j = \inf\{t: d(y_h, Y^T) \leq \varepsilon_0, t > \omega_h^{j-1} + T_b\} \quad (5)$$

with $\tau_h^0 = 0$, $\omega_h^0 = 0$, and $j = 0, 1, \ldots$. Moreover, $T_b$ is the binding time of a transcription factor, $d(\cdot,\cdot)$ is the distance, $\varepsilon_0$ is the minimal distance between gene and factor if they are not bound (in simulation we set $\varepsilon_0 = r_g + r_f$, where $r_g$ and $r_f$ are the radiuses of gene and factor, respectively), and $X^T$ and $Y^T$ are the sets of all available (unbinded) transcription factors at time $t$, i.e.

$$\begin{aligned} X^T &= \{x_i^T | d(x_i^T, \{x_1, \ldots, x_m\}) > \varepsilon_0\}; \\ Y^T &= \{y_i^T | d(y_i^T, \{y_1, \ldots, y_n\}) > \varepsilon_0\}; \end{aligned} \quad (6)$$

Once a transcription factor binds to a gene of the same family, they will move together with gene diffusion coefficient $\sigma^g$ (which is much slower than transcription factor diffusion coefficient $\sigma^T$), i.e.,

$$x_i(t) = x_j^T(t) = X_i(t) \text{ if } d(x_i(\tau_i^k), x_j^T(\tau_i^k)) \leq \varepsilon_0. \quad (7)$$

When the bound gene-factor enters a factory, transcription starts. The transcription time for both X and Y gene are given by

$$\text{X gene}: \quad \xi_{h,s}^j = \inf\{t: d(x_h, F_s) \leq \varepsilon_0, t > \xi_h^{j-1} + T_t\} \quad (8)$$

$$\text{Y gene}: \quad \eta_{h,s}^j = \inf\{t: d(y_h, F_s) \leq \varepsilon_0, t > \eta_h^{j-1} + T_t\} \quad (9)$$

where $j = 0, 1, \ldots$, $\xi_{h,s}^0 = 0$, $\eta_{h,s}^0 = 0$, $F_s$ is the $s$th factory, $s = 1, 2, \ldots, N_f$, and $T_t$ is the transcription time length. We have

$$x_h(t) = x_h(\xi_{h,s}^j), \quad \xi_{h,s}^j < t < \xi_{h,s}^j + T_t\} \quad (10)$$

$$y_h(t) = y_h(\eta_{h,s}^j), \quad \eta_{h,s}^j < t < \eta_{h,s}^j + T_t\} \quad (11)$$

The physical meaning is clear: when the transcription starts, the gene is frozen and stays in the factory. For a given factory $s$, we can calculate the co-localization event. Define the co-localization event as the counting process of the inter co-localization interval $T_{c,xx}$ between one X gene and another X gene (see Fig. 1b) as

$$N_{xx}(s, [0\ T]) = (\text{\# co-localization events in time window } [0\ T]).$$

We can define the co-localization event between X gene and Y gene $N_{xy}(s,[0\ T])$ similarly. Let $u_{h,s}(t)$, $v_{h,s}(t)$ (or $u_{h,s}^T(t)$, $v_{h,s}^T(t)$) be the indicator function of the gene (or factor) transcription event of the $s$th factory for the $h$th gene (X or Y). Note that each process $u_{h,s}(t)$, $v_{h,s}(t)$ (or $u_{h,s}^T(t)$, $v_{h,s}^T(t)$) is a dichotomous random process.

The quantity we intend to calculate is

$$\lim_{T\to\infty} \frac{N_{xx}(s,[0,T])}{N_{xy}(s,[0,T]) + N_{xx}(s,[0,T])}. \quad (12)$$

However, there is a problem if we calculate the ratio as above. When we count the events of $N_{xx}(s,[0\ T])$, the population size is $m(m-1)$, but for $N_{xy}(s,[0\ T])$, it is $mk$. Hence we define

$$r_{xx/xy} = \lim_{T\to\infty} \frac{\frac{N_{xx}(s,[0,T])}{m-1}}{\frac{N_{xy}(s,[0,T])}{k} + \frac{N_{xx}(s,[0,T])}{m-1}}. \quad (13)$$

as the co-localization ratio of X gene (co-localization ratio of Y gene can be similarly defined as $r_{yy/yx}$). When $r_{xx/xy}$ is larger than 0.5, an X gene tends to transcript with another X gene more often in a factory. Let us first confirm that $r_{xx/xy}$ is independent of time and converges to a constant rapidly. From the definition of $N_{xx}(s, [0\ T])$, it is the counting process of a renewal process with the inter co-localization interval $T_{c,xx}$. From the renewal theorem [38,39] we know that when $T\to\infty$,

$$\lim_{T\to\infty} E\left(\frac{N_{xx}}{T}\right) = \frac{1}{E(T_{c,xx})}, \text{ and } \lim_{T\to\infty} E\left(\frac{N_{xy}}{T}\right) = \frac{1}{E(T_{c,xy})}. \quad (14)$$

Hence, as $T$ is large enough we should have

$$r_{xx/xy} = \frac{kE(T_{c,xy})}{(m-1)E(T_{c,xx}) + kE(T_{c,xy})}. \quad (15)$$

Therefore whether there is a co-localization event in the nucleus is completely determined by the inter co-localization interval distribution $T_{c,xx}$ and $T_{c,xy}$.

### Fractional Brownian motion

Is the standard Brownian motion good enough to match the experimental data? Chromatin loci are highly mobile but their motion is restricted within confined volumes. Each gene is constrained by interactions with immobile nuclear structures. Although chromosomes are relatively static, individual chromatin domains undergo Brownian motions and can extend far beyond the edges of their chromosome territory.

A normalized fraction Brownian motion $B_H(t)$ is a continuous-time Gaussian process starting at zero, with mean zero, and having the following covariance function

$$E[B_H(t)B_H(s)] = \frac{1}{2}(|t|^{2H} + |s|^{2H} - |t-s|^{2H}) \quad (16)$$

where $H$, called the Hurst index or Hurst parameter associated to the fractional Brownian motion, is a real number in $[0,1]$.

The value of $H$ determines what kind of process the fraction Brownian motion is:

- if $H = 0.5$, the process is in fact a standard Brownian motion;
- if $0.5 < H < 1$, the increments of the process are positively correlated (super-diffusion);
- if $0 < H < 0.5$, the increments of the process are negatively correlated (sub-diffusion).

According to Eq. (16), when $H$ is greater than 0.5, it moves faster than the normal diffusion ($H = 0.5$), hence it is called superdiffusion. We also use fractional Brownian motion in the model developed in the previous subsection.

### Co-localization moment and ratio between genes

To compare with experimental data, next we introduce some quantities which are experimentally measurable. The transcription rate of an X gene is

$$r_1(u_1) = \lim_{s \to \infty} \frac{\int_0^s (\sum_{j=1}^{N_f} u_{1,j}(t)) dt}{S} = N_f E[u_{1,1}(t)], \quad (17)$$

where $E[\cdot]$ stands for the expectation, and $N_f$ is the number of transcription factories. This means that for an X gene, its transcription rate depends on the number of factories and the probability that this gene is being transcribed over time inside each factory. Similar definition can be given for Y gene transcription rate. The co-localization between two X genes is defined as

$$r_2(u_1, u_2 | u_1) = \lim_{s \to \infty} \frac{\int_0^s (\sum_{j=1}^{N_f} u_{1,j}(t) u_{2,j}(t)) dt}{\int_0^s (\sum_{j=1}^{N_f} u_{1,j}(t)) dt} = \frac{E[u_{1,1}(t) u_{2,1}(t)]}{E[u_{1,1}(t)]}, \quad (18)$$

describing the probability that when an X gene is being transcribed in a factory, another X gene is also being transcribed in the same factory at the same time. Similarly the co-localization ratio for an X and a Y gene is given by

$$r_2(u_1, v_1 | u_1) = \frac{E[u_{1,1}(t) v_{1,1}(t)]}{E[u_{1,1}(t)]}, \quad (19)$$

When X genes and Y genes are independent, we have

$$r_2(u_1, v_1 | u_1) = E[v_{1,1}(t)] = \frac{r_1(v_1)}{N_f} \quad (20)$$

The X-X genes co-localization ratio defined before is simply given by

$$r_{xx/xy} = \frac{r_2(u_1, u_2 | u_1)}{r_2(u_1, v_1 | u_1) + r_2(u_1, u_2 | u_1)} \quad (21)$$

This should give us a clear explanation why we call $r_{xx/xy}$ the co-localization ratio. Therefore, when two X and X genes are colocalized, it should have

$$r_{xx/xy} > 0.5.$$

### Co-localization moment and ratio between genes and factors

Defined

$$r_1(u^T) = \lim_{s \to \infty} \frac{\int_0^S \chi(\sum_{j=1}^{N_f} \sum_{i=1}^{N_T} u_{i,j}^T(t)) dt}{S} = E[\chi(\sum_{j=1}^{N_f} \sum_{i=1}^{N_T} u_{i,j}^T(t))] \quad (22)$$

as the X transcription factor association rate with all factories, where $N_T$ is the number of transcription factors and $\chi(w)$ is the indicator function, i.e.

$$\chi(w) = \begin{cases} 1, & w > 0 \\ 0, & otherwise \end{cases}$$

We assume that all processes are stationary. When $u^T_{i,j}(t)$ is sparse, we have

$$E[\chi(\sum_{j=1}^{N_f} \sum_{i=1}^{N_T} u_{i,j}^T(t))] = E[\sum_{j=1}^{N_f} \sum_{i=1}^{N_T} u_{i,j}^T(t)]. \quad (23)$$

Hence the transcription factor association rate $r_1(u^T)$ is simply $N_f N_T$ $E[u^T_{1,1}(t)]$. The advantage of our approach over the experimental is that we have a dynamical model and we can concentrate on each individual transcription factory. To this end, we will concentrate on the dynamic behaviour of a single transcription factory: peer through one single factory. Under the ergodicity assumption, we intend to match the modelling results with experimental results which are obtained with spatio average. Hence we drop the transcription factory subscript $j$ from now on. In our simulations, for a fixed number of transcription factors, we find a binding time so that X factors (as Klf1 in Schoenfelder et al [7]) are colocalized with the factories with a rate

$$r_1(u^T) = E[\chi(\sum_{i=1}^{N_T} u_i^T(t))]. \quad (24)$$

The colocalizaton ratio between the first X gene and X factors is given by

$$r_2(u_1, u^T | u_1) = \lim_{s \to \infty} \frac{\int_0^s \chi(u_1(t)(\sum_{j=1}^{N_T} u_j^T(t))) dt}{\int_0^s \chi(u_1(t)) dt}$$

$$= \frac{E[\chi(u_1(t)(\sum_{j=1}^{N_T} u_j^T(t)))]}{E[u_1(t)]}. \quad (25)$$

This gives us the Klf1-associated ratio for the first X gene (say, Hbb, Hba, Hmbs and Epb4.9 in Schoenfelder et al [7]) per factory. Again, when the event is sparse, we have

$$r_2(u_1, u^T | u_1) = \frac{E[\chi(u_1(t)(\sum_{i=1}^{N_T} u_i^T(t)))]}{E[u_1(t)]} = \frac{N_T E[u_1(t) u^T_1(t)]}{E[u_1(t)]}. \quad (26)$$

Similarly,

$$r_2(v_1, u^T | v_1) = \frac{E[\chi(v_1(t)(\sum_{i=1}^{N_T} u_i^T(t)))]}{E[v_1(t)]}. \quad (27)$$

is the Klf1-associated rate with Y genes or Z genes. When they are sparse and independent, it equals $\mathcal{N}_T \, E[u^T_{1,1}(t)]$. A pair of X genes colocalized ratio per factory with Klf1 is

$$r_3(u_1, u_2, u^T | u_1, u_2) = \frac{E[\chi(u_1(t)u_2(t) \sum_{i=1}^{N_T} u_i^T(t))]}{E[u_1(t)u_2(t)]}. \quad (28)$$

Since X and Y genes are independent in our model, the co-localization ratio of an X and a Y gene pair with Klf1 becomes

$$r_3(u_1, v_1, u^T | u_1, v_1) = \frac{E[\chi(u_1(t)v_1(t) \sum_{i=1}^{N_T} u_i^T(t))]}{E[u_1(t)v_1(t)]}. \quad (29)$$

When they are sparse and independent, we have

$$r_3(u_1, v_1, u^T | u_1, v_1) = \frac{N_T E[v_1(t)] E[u_1(t) u_{1,1}^T(t)]}{E[v_1(t)] E[u_1(t)]}$$
$$= \frac{N_T E[u_1(t) u_{1,1}^T(t)]}{E[u_1(t)]} = r_2(u_1, u^T | u_1). \quad (30)$$

## Author Contributions

Conceived and designed the experiments: PF JF. Performed the experiments: JK BX YY CH. Analyzed the data: JK BX. Contributed reagents/materials/analysis tools: WL PF JF. Wrote the paper: JK BX PF JF.